# A viscous magnetohydrodynamic Kelvin–Helmholtz instability in the interface of two fluid layer: Part II. An application to the atmosphere of the Sun

Huseyin CAVUS* and G.A. Hoshoudy**

*Canakkale Onsekiz Mart University, Arts & Science Faculty, Physics Department 17100, Canakkale, Turkey, h_cavus@comu.edu.tr

**Department of Applied Mathematics, Faculty of Science, South Valley University, Kena, 83523, Egypt, g_hoshoudy@yahoo.com

**Abstract:** The main aim of this submission is to investigate the effects of some parameters like wave number, shear velocity, magnetic field and temperature for the growth rate of the magnetized Kelvin-Helmholtz instability (KHI) with incessant profiles through interface of two viscous fluid layer occurred in the solar atmosphere using the model of Hoshoudy, Cavus and Mahdy (2019). In this examination, the presence of KHI is identified for the various cases of wave number, magnetic field, shear velocity and temperature in the solar atmosphere. The sensible values of these parameters were acquired.

## 1. Introduction

The instability could be described as an unsteady consequence of a small disturbance. According to Chandrasekhar (1961), any physical case ought to possess a free energy supply to produce instability. A small agitation to the equilibrium is exerted, it will thrive in amplitude to unsteady phases (Lobanov et al., 2003). KHI should be well comprehended for plasma, fluid dynamics uses (Foullon et al., 2011; Ofman and Thomson, 2011). There can be found many theoretical works on this subject. Wang et al. (2009) derived an analytic formula for the linear growth rate and frequency of the KHI in fluids of density gradient in their study of the destabilizing effect of density gradient on the KHI. They found that the density gradient effect enforces the destabilization of the KHI by enhancing its linear growth rate in the direction normal to the perturbed interface. In their other study, Wang et al. (2010) worked on the linear growth rate and the frequency for the combination of KHI and Rayleigh–Taylor instability (RTI) using continuous density and velocity profiles. It was obtained that the density transition layer decreases the linear growth rate in RTI, especially for the short perturbation wavelength. However, the linear growth rate of the KHI increases by the density gradient effect but diminishes by the velocity transition layer. Zhao et al. (2014) studied the effects of a continuous magnetic field pointing in the direction of flow on the noncompressible KHI by obtaining the linear ideal MHD equations. They reached that the frequency of the KHI is not affected by the magnetic field. The magnetic field influence diminishes the linear growth of the KHI, while the magnetic field gradient scale length effect enhances its linear growth. The KHI can even be entirely inhibited when the magnetic field is sufficiently large

Mishin and Tomozov (2016) presented the theoretical results of the KH instability captured by the linear approximation. They showed that the RTI could meaningfully grow the KHI in the higher regions due to interface accelerations or its curvature. Special attention is focused on the compressibility effect on the supersonic shear flow instability in the solar wind and solar CMEs. Shear type flows can be seen in many solar atmospheric structures and are exposed to both fluid and resistive instabilities. These instabilities may play an essential character in the construction and acceleration of both the slow and fast parts of the solar wind (Velli et al., 2003). In Landi and Velli's (2009) work, it was demonstrated that the outcome of KHI type shear-flow instabilities might create much of the monitored events observed in the atmosphere of the Sun such as wind, pressure-balanced structures in the solar wind and coronal plumes. Flows and instabilities take a central position for the MHD plasma of the solar atmosphere (Foullon et al., 2011; Ofman and Thomson, 2011). Observations indicate the existence of governed flows in which the surface and body waves become unstable under certain flow and plasma circumstances. The study of KHI in the atmosphere of the Sun is consequently meaningful.

Effects of viscosity in the solar atmosphere were studied by several papers (Kimura et al., 1998; Mann et al., 1994; Kimura and Mann, 1998; Krivov et al., 1998; Mann et al., 2000). They studied somehow the porous structure of the solar corona and its connection to visible and infrared brightness. They found that the dynamics of near-solar grains is contingent radically on their sizes, chemical composition, number density, temperature and structure in cases of relatively small dielectric grains, may severely be correlated to the solar activity cycle. The diminish of grains' sizes and the dynamics of particles in the orbital plane were well defined by taking into account solar gravity and radiative forces being relatively insensitive to the electromagnetic force. According to their predictions for the sources and transport of dust to the near-solar region, they derived a representative set of trajectories of dust grains by numerical integrations. They obtain the spatial distribution of different dust populations within 10 solar radii from the Sun. For the radial nature, they obtained the dust number density to be enhanced by a factor of 1 to 4 in a typical heliocentric distance zone with a width of 0.2 solar radii in the region of formation of a dust ring depending on the materials and porosities considered.

The KHI is investigated because it is treated as a new potential reason for the development of magnetic reconnection and heating in the atmosphere of the Sun. The leading goal of this paper is to make an investigation of magnetohydrodynamic KHI with continuous profiles through a viscous medium in the solar atmosphere. The model of Hoshoudy, Cavus and Mahdy (2019) will be used as a method. The mathematics used in this work is given in the next section. The consequences of shear velocity, wave number, magnetic field, and temperature on the growth rate of KHI are looked. The physical circumstances and model results are presented in Section 3 for the atmosphere of the Sun. The conclusion is given in Section 4 with some correlations to similar studies.

## 2. Fundamental Physics and Mathematics

We start with equations of momentum and continuity of fluid, where the fluid is permeated by a magnetic field. Under the aforementioned assumptions, equations, that are (Wang et. al., 2009 and 2010; Ye et al., 2011; Hoshoudy, Cavus and Mahdy; 2019)

$$\rho \left( \frac{\partial}{\partial t} + \mathbf{u} \cdot \nabla \right) \mathbf{u} = -\nabla P + \rho \mathbf{g} + \frac{1}{\mu_0} (\nabla \times \mathbf{B}) \times \mathbf{B} - \frac{\rho \nu}{k_I} \mathbf{u} \tag{1}$$

$$\frac{\partial \rho}{\partial t} + \nabla \cdot (\mathbf{u} \rho) = 0, \tag{2}$$

$$\frac{\partial \mathbf{B}}{\partial t} = \nabla \times (\mathbf{u} \times \mathbf{B}), \tag{3}$$

$$\nabla \cdot \mathbf{B} = 0. \tag{4}$$

If fluid elements move without changing density is to say that the Lagrangian total derivative of density is zero, that is,

$$\frac{d\rho}{dt} = \frac{\partial \rho}{\partial t} + \mathbf{u} \cdot \nabla \rho = 0. \tag{5}$$

Where $\mathbf{u}$ is the velocity, $\rho$ is the density, $p$ is the pressure, $\mathbf{g}$ is the gravitational acceleration, $\nu$ is the kinematic viscosity, $k_I$ is the intrinsic permeability, $\mathbf{B}$ is the magnetic field and $\mu_0$ is magnetic permeability. The MHD equations to be worked out are composed of the momentum transfer equation (1), the equation of continuity (2). Induction, magnetic monopole and the incompressibility flow are expressed by the equations (3)-(5) respectively. These equations are nonlinear and analytical solutions are difficult to obtain. We shall continue as in Hoshoudy, Cavus and Mahdy (2019). In our study, we consider the motion in two-dimensional $(x, y)$, where the two-dimensional disturbances are more unstable than three-dimensional which is proved by Squire's theorem (Todd et al., 1986). In addition, we take into account that, the initial velocity in $x$ – direction, the initial pressure and the initial magnetic field in $y$ – direction, i.e.

$$\mathbf{u}^{(1)}(x,y,t) = \left( u_x^{(1)}, u_y^{(1)}, 0 \right), \ p^{(1)}(x,y) = p^{(1)}(x,y,0), \ \mathbf{B}^{(1)}(x,y) = \left( B_x^{(1)}, B_y^{(1)}, 0 \right). \tag{6}$$

Now, we give a small perturbation to the system, where the perturbations in the velocity, pressure, density and the magnetic field, respectively, are

$$\mathbf{u} = \mathbf{U}^{(0)} + \mathbf{u}^{(1)}, \ p = p^{(0)} + p^{(1)}, \ \rho = \rho^{(0)} + \rho^{(1)} \ \text{and} \ \mathbf{B} = \mathbf{B}^{(0)} + \mathbf{B}^{(1)}. \tag{7}$$

In addition, we take the initial velocity, magnetic field, density and pressure are a function in the vertical axis, i.e.

$$\mathbf{U}^{(0)} = \left( u_x^{(0)}(y), 0, 0 \right), \ \mathbf{B}^{(0)} = \left( B_x^{(0)}(y), 0, 0 \right), \ \rho^{(0)} = \rho^{(0)}(y), \ p^{(0)} = p^{(0)}(y) \ . \tag{8}$$

Assuming small perturbations, in the linearization process, we can drop second and higher order terms in these perturbations. We consider the perturbation in quantities $\rho^{(1)}$, $\mathbf{u}^{(1)}(u_x^{(1)}, u_y^{(1)})$, $p^{(1)}$ and $\mathbf{B}^{(1)} = (B_x^{(1)}, B_y^{(1)})$ $p^{(1)}$ as

$$\left\{ \rho^{(1)}, \mathbf{u}^{(1)}, p^{(1)}, \mathbf{B}^{(1)} \right\} = \left\{ \rho(y), \mathbf{u}(y), p(y), \mathbf{B}(y) \right\} \exp \left\{ \mathrm{i}kx + \sigma t \right\} \tag{9}$$

in which, $\sigma = \gamma - \mathrm{i}\omega$ ($\gamma$ and $\omega$ are the linear growth rate and the frequency of perturbations). Using these, by following the similar method in Hoshoudy, Cavus and Mahdy (2019) the general characteristic equation for the instability growth rate is obtained as (see Eq. 20 of this paper),

$$\gamma^2 + \left(\frac{\nu}{k_I}\right)\gamma - \left(U^{(2)} - U^{(1)}\right)^2 k^2 \left\{ \begin{array}{c} \dfrac{A^2}{2(1+1/kL_\rho)} - \dfrac{(1-A^2)}{2(1+1/kL_u)} - \dfrac{A^2}{2(1+1/kL_\rho+1/kL_u)} - \\ \dfrac{A^2}{4}\left[\dfrac{1}{(1+1/kL_\rho)} + \dfrac{1}{(1+1/kL_u)} - \dfrac{1}{(1+1/kL_\rho+1/kL_u)}\right]^2 + \\ \dfrac{1}{4(1+2/kL_u)} + \dfrac{(1-A^2)}{4} \end{array} \right\}$$

$$\frac{gAk}{(1+kL_d)} + \frac{k^2}{\mu_0(\rho_+ + \rho_-)}\left\{\left(B^{(1)}\right)^2 + \left(B^{(2)}\right)^2 + \frac{\left(B^{(1)} - B^{(2)}\right)^2}{2+4/kL_B} - \frac{\left(B^{(1)} - B^{(2)}\right)^2}{1+1/kL_B}\right\} = 0 \qquad (10)$$

$L_\rho$, $L_u$ and $L_B$ are the density, velocity and magnetic field gradients scale lengths, respectively. $U^{(2)}$ ($U^{(1)}$) is velocity away from the interface in $x$–direction of the upper (lower) fluid, $B^{(2)}$ ($B^{(1)}$) is the magnetic field of the upper (lower) fluid and $A = \dfrac{\rho_+ - \rho_-}{\rho_+ + \rho_-}$ is Atwood number, where $\rho_+$ ($\rho_-$) is the density away from the interface of the upper (lower) fluid.

In particular, if $\gamma$ is real positive, Eq. (9) simply represents an instability so that the perturbation grows exponentially with the time that is the system is unstable, otherwise, the solution is oscillatory. In the rest of the paper, the investigation of KHI will be studied by $\gamma$ acquired from Equation (10) for the solar atmosphere by the use of three-dimensional graphs and contour plots.

## 3. Results and their analysis

### 3.1. Physical parameters in the atmosphere of the Sun

We calculate the solutions to Eq. (10) for the atmosphere of the Sun by formulas given in Section 2 using an algorithm developed and adapted to Maple 15. We used the values of physical parameters taken from previous papers related to the atmosphere of the Sun as listed in Table 1.

| Parameter | Symbol | Value |
|---|---|---|
| Gravitational acceleration | $g$ | 140 m/s$^2$ |
| Plasma densities | $\rho_1$ | $\rho_1 \approx 8.5\times10^{-11}$ m$^{-3}$ |
|  | $\rho_2$ | $\rho_2 \approx 1.7\times10^{-12}$ kg m$^{-3}$ |
| Magnetic fields | $B_1$, $B_2$ | 4 G, 5 G, 8 G |
| Shear velocity | $\Delta U = U^{(2)} - U^{(1)}$ | 600–900 km s$^{-1}$ |
| Intrinsic permeability | $k_1$ | 0.1 m$^2$ (no porosity means $k_1 \to \infty$) |
| Density, velocity and magnetic gradients scale lengths | $L_V$, $L_\rho$ and $L_B$ | All of them are taken as 10$^{-6}$ m (theoretically 0) |
| Kinematic viscosity | $\nu$ | $2.2\times10^{-20}\,T^{5/2}$ m$^2$ s$^{-1}$ |
| Temperature | $T$ | $0.5\times10^6$ K |
| Wave number | $k$ | $10^8$ m$^{-1}$ |

**Table 1.** Physical parameter in solar atmosphere (Priest, 1983; Hood, 1992; Nakarikov et al., 1996; Gombosi, 1998; Choudhuri, 1998; Belien et al., 1999; Keppens and Goedbloed, 2000; Velli et al., 2003; Zaliznyak et al., 2003; Verdini et al., 2005; Selwa et al., 2005; Zhelyazkov, 2009; Rathinavelu et al., 2009; Wang et al., 2009; Wang et al., 2010; Ryutova et al., 2010; Foullon et al., 2011; Ofman and Thomson, 2011; Cavus and Kazkapan, 2013).

### 3.2. Results

Operating the parameters given in Table 1, it is attempted to solve Equation (10) for $\gamma$ beneficial to investigate the KHI in the atmosphere of the Sun. In Figures 1 – 8, the changes of $\gamma$ are depicted according to $B$–field, $\Delta U$, $T$ and $k$.

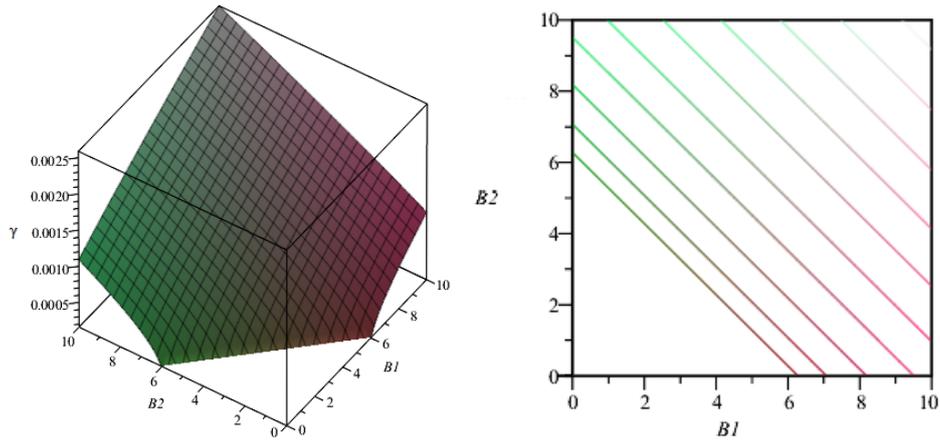

Figure 1. Change of growth rate (in s$^{-1}$) with respect to $B_1$ and $B_2$ (both are in G) for $\Delta U$=600 km s$^{-1}$

In Figure 1, the three-dimensional graph and contour plot of growth rate is given with respect to $B_1$ and $B_2$ for the shear velocity value of $\Delta U$ = 600 km s$^{-1}$. The system becomes stable, the solution is being oscillatory in the triangular region defined by $B_1$ and $B_2$ are less than 6.2 G. The maximum value of the growth rate is greater than 0.0025 for $B_1 = B_2 = 10$ G. Figure 2 is drawn for $\Delta U$ =900 km s$^{-1}$. In this case, it is stable in the triangular region for both $B_1$ and $B_2$ < 9 G. The growth rate's maximum value is much greater than 0.002 for $B_1$ and $B_2$ are equal to 10 G.

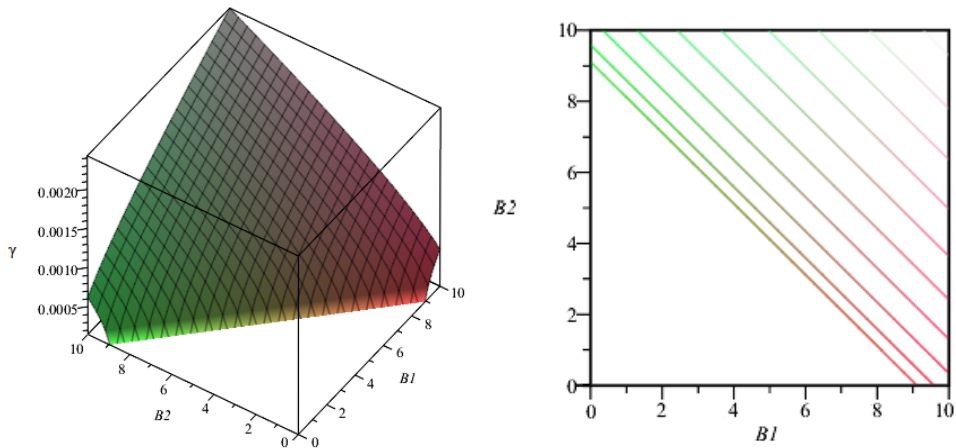

Figure 2. Same as Figure 1 but for $\Delta U$ = 900 km s$^{-1}$

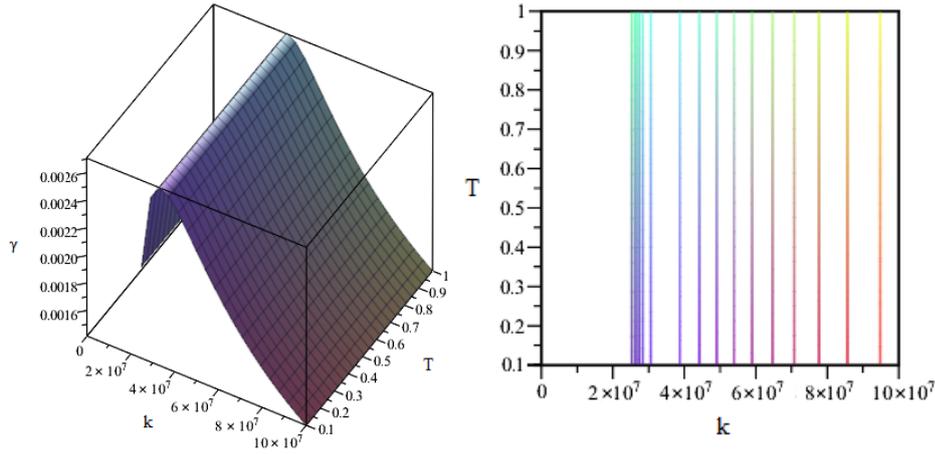

Figure 3. Variation of growth rate with respect to $k$ (in m$^{-1}$) and $T$ (in MK) for $B_1 \neq B_2$ and $\Delta U = 600$ km s$^{-1}$.

In Figure 3, the change of $\gamma$ is given with respect to $k$ and $T$ for $B_1 \neq B_2$ (i.e. $B_1 = 4$ G, $B_2 = 8$ G) and $\Delta U = 600$ km s$^{-1}$, it is unstable for all $T$. However, it is stable for wave number $k < 2.2 \times 10^7$ m$^{-1}$. The maximum value of $\gamma$ reaches the values greater than 0.0026 at $k = 3.6 \times 10^7$. The $\Delta U = 900$ km s$^{-1}$ case is drawn in Figure 4. The stable oscillatory solution occurs for the wave number values less than $5.5 \times 10^7$ m$^{-1}$. The maximum value of $\gamma$ is greater than 0.001 at $k = 8 \times 10^7$ m$^{-1}$.

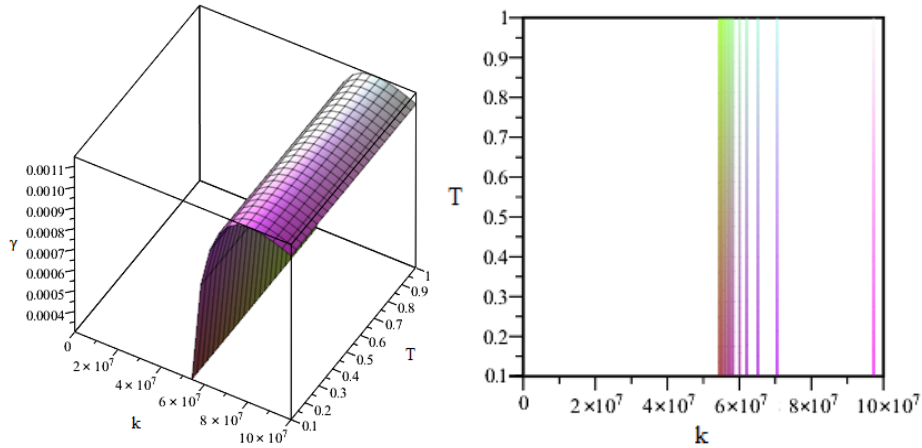

Figure 4. Same as Figure 3 but for $\Delta U = 900$ km s$^{-1}$.

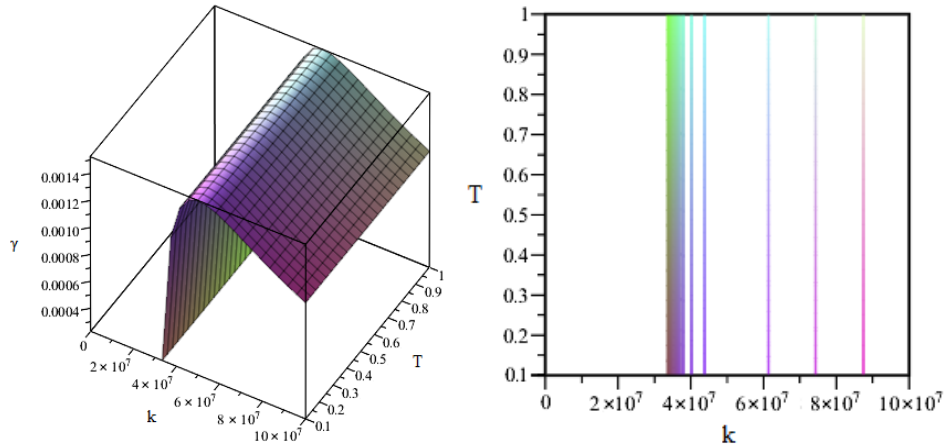

Figure 5. Same as Figure 3 but for $B_1 = B_2$

In Figure 5 and 6, the three-dimensional graph and contour plot of growth rate is given with respect to $k$ and $T$ for $B_1 = B_2 = 5$ G for $\Delta U = 600$ km s$^{-1}$ and $\Delta U = 900$ km s$^{-1}$ respectively. They are again unstable for all $T$. For $\Delta U = 600$ km s$^{-1}$, the oscillatory solution occurs for $k < 3.4 \times 10^7$ m$^{-1}$ as shown in Figure 5. The maximum values of the growth rate, greater than 0.0014, occurred at $k = 5 \times 10^7$ m$^{-1}$. Figure 6 is drawn for $\Delta U = 900$ km s$^{-1}$. The system is stable for $k < 7.9 \times 10^7$ m$^{-1}$. Growth rate's maximum value is about 0.006 for $k = 10 \times 10^7$ m$^{-1}$.

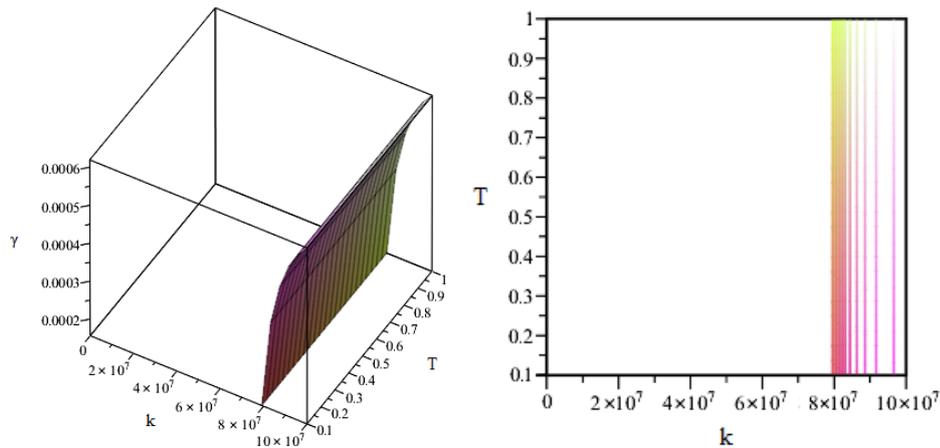

Figure 6. Same as Figure 5 but for $\Delta U = 900$ km s$^{-1}$

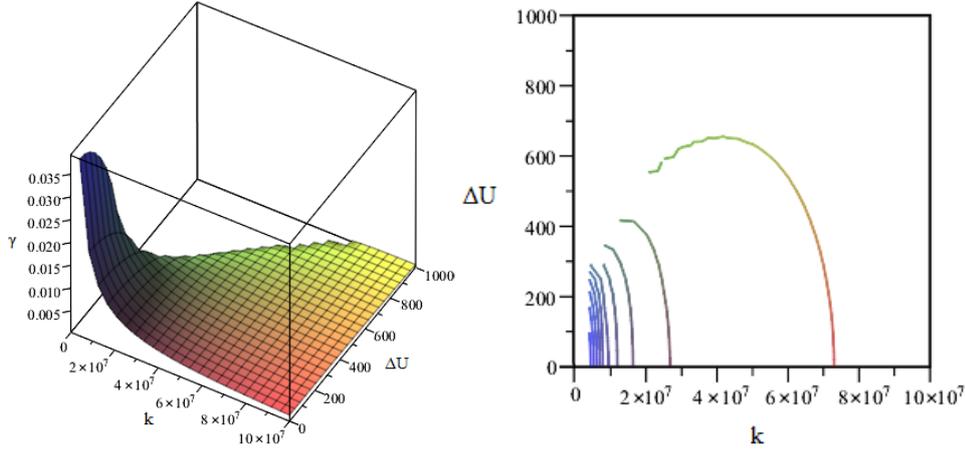

Figure 7. The plot of growth rate according to $k$ (in m$^{-1}$) and $\Delta U$ (in km s$^{-1}$) for $B_1 \neq B_2$.

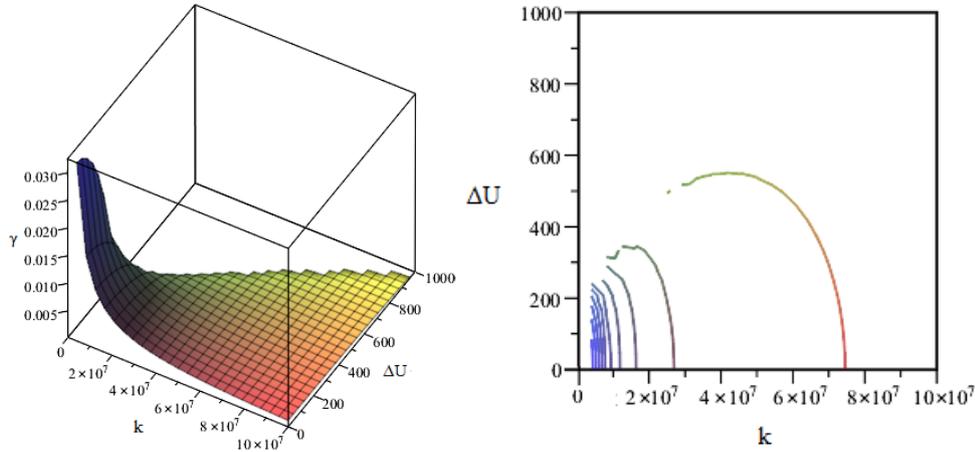

Figure 8. Same as Figure 7 but for $B_1 = B_2$.

In Figures 7 and 8, the variations of γ are drawn according to the wave number and shear velocity for $B_1 \neq B_2$ and $B_1 = B_2$ respectively. In both case the system; tend to be more unstable for the smaller values of $k$ and $\Delta U$. Especially as shown in the contour plot the instability occurs at the left part for all values of $k$. The maximum values of γ are about 0.035 and 0.030 respectively.

## 4. Discussion and Conclusion

For KHI instability, which is so essential in cosmological plasmas, it has not been available to have confirmation for the evolution of this instability in the atmosphere of the Sun as said in Foullon et al. (2011) and Cavus and Kazkapan (2013). New facilities for observing the Sun permit to catch and recognize KHI type formations, which can be regarded as a new reason for the development of the heating and magnetic reconnection in the solar atmosphere at which the instability has occurred.

In present work, the fundamental goal is to investigate the influences of $B$–field, $\Delta U$, $T$, $k_1$ and $k$ on the $\gamma$ of the magnetized KHI with continuous profiles through a viscous medium in the solar atmosphere using the model of Hoshoudy, Cavus and Mahdy(2019). In this investigation, decisive values of $B$–field, $\Delta U$, $T$ and $k$ in the solar atmosphere captured from the graphs. This study can be considered as a first KHI study in which the porosity was taken into consideration for the solar atmosphere.

We first found that KHI might occur all temperature range depending on other physical parameters as shown in figures. Using the intrinsic permeability, $k_1 = 0.1$ m$^2$, (used as a scale for viscosity and porosity) the critical values of $B$–field were obtained about 6.2 G and 9 G respectively for $\Delta U$ of 600 km s$^{-1}$ and 900 km s$^{-1}$. These are in the adequate values for $B$–field in the atmosphere of the Sun as said by Priest (1983 and 2000). Here, the received value of decisive values of shear velocity are about 560 km s$^{-1}$ and 660 km s$^{-1}$ for different values of $B$–fields $T$ as shown in contour plots of Figure 7 and 8. These value are in good correlation with the explanations fast type of solar winds as stated in Priest (1983 and 2000) and Gombosi (1998). As stated in the works of Knoll and Brackbill (2002) and Lapenta and Knoll (2003), obtained $\Delta U$ (i.e. about 600 km s$^{-1}$) are in super-Alfvenic regime KHI is expected to occur.

Obtained $\gamma$ of instability increases with the values shear velocity, it is necessary to have bigger values of $B$–field necessary to stabilize as mentioned in Lapenta and Knoll (2003). We saw that the $B$–field parallel the shear, which is along with the interface, could have to stabilize influence (Ofman and Thompson, 2011; Lysak and Song, 1996). We found that the strong magnetic field attains intensity as $B$–field becomes enough to stabilize as it is stated in the work of Zhao et al. (2014). Plasma escape occurs in areas formed by KH vortex. We believe that the mechanism offered here guarantees that further studies will be possible. A clear understanding of the KHI's function in reconnection needs a complete three-dimensional modelling of the flows